# Analysis of Two-Tier LTE network with Randomized Resource Allocation and Proactive Offloading

Katerina Smiljkovikj, Aleksandar Ichkov, Marko Angjelicinoski, Vladimir Atanasovski and Liljana Gavrilovska

Ss. Cyril and Methodius University in Skopje, Macedonia
Faculty of Electrical Engineering and Information Technologies
{katerina, ichkov, markoang, vladimir, liljana}@feit.ukim.edu.mk

*Abstract*— The heterogeneity in cellular networks that comprise multiple base stations' types imposes new challenges in network planning and deployment. The Radio Resource Management (RRM) techniques, such as dynamic sharing of the available resources and advanced user association strategies determine the overall network capacity and the network/spectrum efficiency. This paper evaluates the downlink performance of a two-tier heterogeneous LTE network (consisting of macro and femto tiers) in terms of rate distribution, i.e. the percentage of users that achieve certain rate in the system. The paper specifically addresses: (1) the femto tier RRM by randomization of the allocated resources; (2) the user association process by introducing novel proactive offloading scheme and (3) femto tier access control. System level simulation results show that an optimal RRM strategy can be designed for different scenarios (e.g. congested and uncongested networks). The proposed proactive offloading scheme in the association phase improves the performance of congested networks by efficiently utilizing the available femto tier resources. Finally, the introduced hybrid access in femto tier is shown to perform nearly identical as the open access.

I. INTRODUCTION

The increasing need for capacity in cellular networks leads to network densification, which is one of the evolution directions for 5G networks [1]. The densification can be achieved either in space, by increasing the number of network nodes in the system, or in frequency, by utilizing different portions of spectrum in different bands.

The densification by adding new Macro Base Stations (MBSs) is an expensive solution. Additionally, the MBSs are not able to solve the problem of indoor coverage and high data rate for indoor users. Femtocells have emerged as a promising solution since the user-centric deployment of Femto Base Stations (FBSs) is inexpensive and uncoordinated. This makes them preferable for coverage and data rate improvement [2][3]. Network densification by adding additional femto tier of base stations that differs from the macro tier in terms of transmission power, capacity and base station spatial density, results in network heterogeneity.

The Radio Resource Management (RRM) of the femto tier is essential for overall network performance. Unlike the macro tier that is a subject to frequency planning prior to actual deployment, the femto tier is usually deployed sporadically and randomly without any spatial or frequency planning [4].

Additionally, the deployment of the FBSs is usually uncoordinated with the macro tier, which further complicates the design of intelligent strategies for resource allocation and sharing.

Traditional user-to-BS association strategies are mostly BS coverage based and favor the MBSs for user association. In such cases, high portions of the femto layer resources might remain underutilized resulting in degradations of the overall network spectrum efficiency of the. Existing load balancing and offloading schemes reactively address these issues, but they are often time-consuming.

Furthermore, a femto BS can be configured to allow either open, closed or hybrid access to its potential users. Open access allows an arbitrary nearby cellular user to use the femto BS, while closed access restricts only authorized users to connect. Hybrid access can be used to compromise between authorized and non-authorized users. The implementation of access control mechanisms additionally affects the utilization of the network resources.

The analysis of heterogeneous cellular networks requires new metrics for performance evaluation since the main challenge is to maximize network capacity, not coverage [5]. A possible metric under these circumstances is the *rate distribution* defined as the probability that a typical user in the network receives data rate beyond a predefined threshold. Unlike traditional network coverage based metrics, the rate distribution unambiguously captures the aspect of spectrum utilization efficiency in heterogeneous networks.

This paper evaluates the performance of a two-tier LTE network that employs randomized RRM on the femto tier. We introduce novel two-step macro-to-femto offloading scheme, which is used proactively in the user association phase as a viable tool for improvement of the data rate distribution in congested network. Also, the newly introduced hybrid access control on femto tier efficiently utilizes the network resources, while guaranteeing predefined rates to authorized users. Obtained results pinpoint directions for future design of optimal resource sharing and utilization strategies in heterogeneous cellular networks.

The paper is organized as follows. Section II describes the related work in the area. Section III describes the general system model, whereas section IV overviews existing user association strategies and introduces the novel two-step offloading scheme and the femto access control. Section V evaluates the system performance through system level simulation. Finally, section VI concludes the paper.

## II. RELATED WORK

Heterogeneous cellular networks are an area of extensive research recently. Ref. [5] shows that the maximum Signal-to-Interference-and-Noise Ratio (SINR) association is suboptimal in this case, because it leaves unutilized resources in base stations with smaller coverage. Authors in [6] analyze $K$-tier downlink heterogeneous cellular network and derive closed form expressions for coverage probability and average rate. In [7], the authors perform load-aware downlink modeling and the authors in [8] deal with interference management. Other analysis for load distribution in heterogeneous cellular networks is given in [9]. Intelligent interference cancellation technique is proposed in [10], which allows spectrum reuse on the femto tier in the network.

Regarding association, ref. [11] analyzes a class of association metrics, named stationary associations in heterogeneous cellular networks. Ref. [12] analyzes rate distribution under generalized cell selection assuming that shadowing impacts cell selection while fading does not. In [12], a joint algorithm for user association and base station operation is proposed, which allows certain BSs to be turned on/off according to the association metric. Traffic offloading between different tiers is analyzed in [14] and [15].

Unlike previous work [6]-[15], where the models cannot be directly applied in OFDMA-based cellular networks, the analyses in this paper are performed assuming realistic LTE system [16] with predefined minimal amount of frequency resources (the PRBs) that can be assigned to a user that requires service. Additionally, the paper introduces resources randomization in the femto tier and proactive two-step macro-to-femto offloading scheme accompanied with different femto access control schemes.

## III. RADIO RESOURCE MANAGEMENT IN TWO-TIER LTE NETWORK

This section describes the topological organization of the two-tier LTE network, comprising macro and femto tiers, the RRM strategies and the general users' resource allocation adopted by both tiers.

### A. System topology and user distribution

The MBSs are distributed over a specific area, denoted with $A$ in $m^2$ (Fig. 1). The model in this paper adopts regular MBSs distribution, in a grid. Each MBS is at the center of hexagonal cell so that the distance between two adjacent BSs is $d$. The set of MBSs is denoted with $M$ and the number of MBSs is denoted with $N_m$, i.e. $|M|=N_m$. It is further assumed that the downlink transmit power of each MBS is $P_m$ over all available spectrum resources.

The FBSs are randomly distributed over the same area $A$, uncoordinated with the MBSs. The basic difference between the macro and the femto tiers are the transmit powers of the respective BSs. In particular, the downlink transmit power of the FBSs is much lower than the transmit power of the MBSs, $P_f \ll P_m$. Poisson Point Process (PPP) models the randomized and uncoordinated deployment of FBSs, as it is the most frequently used statistical distribution for modeling stochastic, two-dimensional point processes [14]. The intensity of the femto tier PPP is denoted as $\lambda_f$ in $m^{-2}$. The average number of FBSs in the area is $N_f = \lambda_f A$. The set of all FBSs is denoted as $F$.

The users are assumed to be distributed according to PPP (Fig.1), with intensity $\lambda_u$ with average number $N_u = \lambda_u A$. The set of user is denoted with $U$.

### B. Radio Resource Management

The radio access technology in focus is LTE's OFDMA with total system bandwidth $W$ in $MHz$. The smallest resource unit that can be allocated to a single user is referred as Physical Resource Block (PRB) with $W_{PRB} = 180 kHz$ of bandwidth in frequency domain and 2 slots in time domain. Thus, the total bandwidth of the system can be represented as the total number of PRBs available to the system.

The macro tier uses Hard Frequency Reuse (HFR) with Frequency Reuse Factor (FRF) $K=3$, dividing the available bandwidth on 3 equal continuous frequency fragments. (Fig. 1). Each MBS gets one fragment, $N_{PRB,m} = N_{PRB}/K$ PRBs and thus, a maximum of $N_{PRB,m}$ users can be associated with a single MBS for downlink transmission. The use of HRF allows to take into account the co-channel interference in downlink from MBSs using the same frequency bandwidth.

The femto tier employs uncoordinated approach for resource allocation in order to mitigate the interference between different FBSs and to alleviate the interference between the macro and the femto tiers. In particular, the femto tier divides the available system bandwidth in several continuous fragments of PRBs in the frequency. The number of fragments is denoted with $n_f$ and each fragment consists of $N_{PRB,f} = N_{PRB}/n_f$ PRBs. Then, each FBS randomly chooses a single fragment. Depending on the overall network traffic load, the femto tier can dynamically adapt its fragment size to varying network conditions. Such femto tier dynamism requires minor coordination with the macro tier to determine the traffic state of the network, which can be easily achieved through the existing and standardized network interfaces.

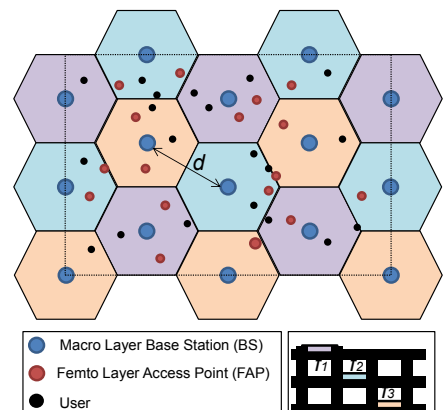

Fig. 1. System topology and macro layer frequency allocation in the observed area $A$

## C. User resource allocation

Generally, the user resource allocation requires each BS to distribute the available physical resources fairly, using equal power allocation to each PRB so that each user gets the maximal possible data rate with respect to the network settings and traffic load. When the number of associated users to BS $i$ is $N_{u,i} < N_{PRB,m}$ if $i \in M$ or, equivalently, $N_{u,i} < N_{PRB,f}$ and if $N_{PRB}$ or $N_{PRB,f}$ are not divisible with $N_{u,i}$, then the remaining PRBs are randomly distributed to the associated users for maximum resource allocation fairness. Note that, when $N_{u,i} = N_{PRB,m}$ for $i \in M$ or $N_{u,i} = N_{PRB,f}$ for $i \in F$, each associated user is allocated a single PRB. The users' resource allocation depends on the user-to-BS association process and the access control mechanisms embedded on the associated BS, described in details in Section IV.

## IV. USER-TO-BS ASSOCIATION FOR DOWNLINK TRANSMISSION AND ACCESS CONTROL

This section provides brief overview of the existing user association strategies, discusses the possibilities for spectrally efficient associations in two-tier network architectures and proposes novel, two-step macro-to-femto offloading scheme that offloads user traffic in the initial user association phase. It also addresses the access control embedded on the femto tier.

### A. User-to-BS association

The user-to-BS association strategy determines which BS the user associates with when the system has data to transmit to the user (downlink user association). The common approach is each user to associate to the BS (MBS or FBS) that maximizes a predefined association metric. Let $A = M \cup F$ denote the set of all BSs in the two-tier network. Then, each user associates with the BS $k$ in accordance with the following, general association rule:

$$k = \arg\max_{i \in A} \{T_i Z_i^{-\gamma}\} \quad (1)$$

In (1), $Z_i$ is the distance between the $i$-th BS and the user, $\gamma$ is the path loss exponent and $T_i$ is referred as association weight. Note that, if $T_i \gg T_j$, $i \in F, j \in M$, then more traffic is routed through the femto tier. Thus, by adjusting the value of the association weight, the system can control how the traffic is distributed among the tiers, even among the different nodes of the tiers and efficient traffic load balancing schemes can be developed. This paper considers several special cases for the association approaches depending on the value of $T_i$:

- *Nearest BS association* (MBS or FBS): $T_i = 1, \forall i \in A$
- *Cell range modification*: $T_i = P_i B_i$ and the range is extended if the bias factor $B_i > 1$ or reduced if $B_i < 1$
- *Femtocell range extension*: if the bias factor $B_i > 1, \forall i \in F$ and $B_i = 1, \forall i \in M$
- *Maximum received power association*: if the bias factor $B_i = 1, \forall i \in A$ then $T_i = P_i, \forall i \in A$

If the number of users that want to associate to a given BS is higher than the number of available PRBs, i.e. $N_{PRB,m}$ for MBS and $N_{PRB,f}$ for FBS, then the system associates the $N_{PRB,m}$ or $N_{PRB,f}$ users with highest values according to the association rule (1). Thus, the scheduling is opportunistic in the frequency domain. The remaining users are dropped and rescheduled for later transmission in subsequent time slots.

In most cases, the general user association rule (1) results in high number of users being associated to the MBSs. This situation is critical in congested networks, in which case there will be high number of denied users by the macro tier and high portions of the physical resources of the femto tier unused. Forcing more traffic routing to the femto tier might also result in congested FBSs.

### B. Two-step, macro-to-femto association algorithm

As a compromising solution, this paper proposes simple, two-step offloading scheme for user association as a compromising solution of the aforesaid issues regarding the user traffic balancing and efficient utilization of the physical resources:

| **Algorithm**: Two-step, macro-to-femto offloading scheme |
|---|
| 1: Number of users that send association requests to MBS $i$ using (1) $\rightarrow N_{u,i}$ |
| 2: Step 1: *Macro layer user association* |
| 3:     If $N_{u,i} \leq N_{PRB,m} \rightarrow$ associate all users with MBS $i$ |
| 4:     Else If $N_{u,i} > N_{PRB,m}$ |
| 5:         Step 2: *Femto layer offloading* |
| 6:         $\rightarrow$ Associate the best $N_{PRB,m}$ to the MBS $i$ |
| 7:         $\rightarrow$ Forward the rest $N = N_{u,i} - N_{PRB,m}$ users to their closest FBSs and use (1) for association |

The basic idea is to allow each user, initially to send association request to the desired BS in accordance with the association rule (1). However, if the base station does not have free PRBs to allocate all tagged users for downlink transmission, the remaining users are not dropped. Instead, they are forwarded to the femto tier to perform association with their respective, closest FBSs. Existing solutions for load balancing, reactively offload the user traffic between the tiers but often they are time consuming leading to suboptimal and inefficient resource utilization. Note that in the above algorithm, the offloading is performed in the association phase and can be regarded as proactive load balancing scheme that tries to avoid congestion in the initial, communication establishment phase. Additionally, the algorithm avoids time and resource consuming re-associations of the dropped users by forwarding them to the femto tier. Due to its simplicity and lack of control information overhead, the proposed algorithm is well suited for scenarios with high number of user association requests at a time, such as public gatherings.

## C. Access control on femto tier

In order to describe the access control procedures, it is important to note that users can be classified into two categories, depending on the connectivity rights that they are given, i.e. subscribers and non-subscribers. A *subscribers* of a femtocell is a user registered in it, while a *non-subscriber* us a user not registered in the femtocell. The subscribers list is decided by the femtocell owners and set up by the operator. From access methods viewpoint, femtocells can be configured to be either open, closed or hybrid. Closed access restricts access only to its subscribers rejecting non-subscribers' requests. If there are no subscribers' requests on the femtocell, its resources remain unused. All users are allowed to connect to an open access femtocell. Therefore, there is no distinction between subscribers and non-subscribers, referred to as users in this case. Combining the advantages of both mechanisms, hybrid access allows preferential access to the subscribers of a femtocell, with higher priority than non-subscribers. If there are non-subscribers' requests, the femtocell allows access to non-subscribers utilizing all of its resources among them. In this sense, hybrid access enhances resource utilization in comparison to closed access scenario, keeping the guaranteed access for the subscribers.

## V. PERFORMANCE EVALUATION

This section focuses on the system level performance analyzing the rate distribution of the previously elaborated two-tier LTE network.

### A. Simulation setup

The system is simulated in MATLAB in order to evaluate the performance of the two-tier cellular network with realistic LTE settings [15]. The simulation parameters are given in Table I.

TABLE I. SIMULATION PARAMETERS

| Simulation setup | |
|---|---|
| *Simulation environment* | MATLAB |
| *Observed area (A)* | 25x25km |
| *Total system bandwidth* | 15MHz |
| *Macro BS transmit power* | 43dBm |
| *Femto BS transmit power* | 20dBm |
| *Distance between two macro BSs* | 5km |
| *Macro frequency reuse factor* | 3 |
| *Number of fragments for femto tier bandwidth randomization* | 1, 3, 5, 15 and 25 |
| *Wireless channel gain distribution* | Rayleigh (unit var.) |
| *Path loss exponent* | 2.3 |
| *Noise power* | $10^{-12}$ |
| *Radius of femtocell coverage* | 18m |
| *Number of subscribers on femtocell* | 3 |

The number of MBSs over the targeted area is 33 and is sufficient for including the effect of macro tier frequency reuse in the system model. The system uses 15MHz of bandwidth corresponding to 75 PRBs. At the macro tier, the total bandwidth is divided in three sub-bands, each containing 25 PRBs. The sub-bands are regularly assigned to the MBSs with $K=3$. At the femto tier, the whole bandwidth is divided into 1, 3, 5, 15 or 25 fragments, each containing 75, 25, 15, 5 or 3 PRBs, respectively. One fragment consists only of frequency continuous PRBs. The FBSs randomly choose one fragment of the pool of possible fragments.

Standard SINR calculation formulae, assuming complete spectrum sharing and reuse, are not applicable in the above described system. The SINR calculation in OFDMA based systems requires incorporation of the number of scheduled PRBs for a particular user and the number of overlapping PRBs with other MBSs or FBSs in the area. Therefore, this paper introduces SINR calculation formula that uses novel approach to calculate the interference from surrounding BSs with overlapping PRBs. For a particular user $i \in U$, the SINR is calculated as:

$$SINR_{ij} = \frac{\frac{\alpha_{ij}P_j}{N_{PRB,j}}\|h_{ij}\|^2 \|x_{ij}\|^{-\alpha}}{\sum_{\substack{m=1 \\ m \neq j \\ j \in M}}^{N_m} \frac{\beta_{im}P_m}{N_{PRB,m}}\|h_{im}\|^2 \|x_{im}\|^{-\alpha} + \sum_{\substack{f=1 \\ f \neq j \\ j \in M}}^{N_f} \frac{\beta_{if}P_f}{N_{PRB,f}}\|h_{if}\|^2 \|x_{if}\|^{-\alpha} + N_0} \quad (2)$$

Equation (2) represents the received SINR at the $i^{th}$ user that is associated to the $j^{th}$ BS ($j$ is either in $M$ or $F$). $\alpha_{ij}$ is random variable that represents the number of PRBs allocated to the user. The total transmit power from the $j^{th}$ base station to the $i^{th}$ user is $\alpha_{ij}P_j/N_{PRB,j}$, where $P_j/N_{PRB,j}$ is the transmit power on one RB from the $j^{th}$ base station. $h_{ij}$ and $x_{ij}$ are the channel fading and the distance between the $i^{th}$ user and the $j^{th}$ BS, respectively. The first sum in the denominator denotes the interference from MBSs in the system, while the second sum denotes the interference from FBSs in the system. The random variables $\beta_{im}$ and $\beta_{if}$ represent the number of overlapping PRBs between the $i^{th}$ user and the interfering BSs, both macro and femto, respectively. The parameter $N_0$ is the noise power.

The rate for the $i^{th}$ user in the system, knowing the received SINR, is calculated as:

$$R_{ij} = \alpha_{ij}W_{PRB}\log_2(1+SINR_{ij}) \quad (3)$$

The rate distribution, or equivalently the probability that certain percentage of users achieve rate higher than a predefined threshold is defined as:

$$\Psi = \Pr[R > \delta \mid R > 0] \quad (4)$$

The system model allows for service denial and user rescheduling in subsequent time slots. However, the rate distribution defined with (4) refers to the rate for the associated users only. The goal is to maximize the resource utilization for the associated users, i.e. to maximize the average rate in the system that can be guaranteed to any associated user.

Dispersing the users into the two categories, the subscribers list of a closed/hybrid access femtocell consists of three randomly chosen users from the femtocell's coverage area within the 18m radius. The femtocell subscribers can not overload the femtocell due to the fact that in worst-case scenario it has three available PRBs ($n_f$ =25). Using the rate distribution (4) referring to all users, we are able to see the impact of the rejected non-subscribers' requests in the system as well as the increased rates per subscribers, considering the embedded access control mechanism.

*B. Simulation results*

The rate distribution of the system for different number of femto tier spectrum fragments and different number of FBSs is shown in Fig. 2. The nearest BS association metric is used to evaluate the performance of the femto tier RRM. The average number of users over the area is $N_u = 10000$. Fig. 2a shows the rate distribution when the average number of FBSs in the network is $N_f = 500$. This scenario corresponds to congested network since the number of available PRBs and the number of downlink user association requests in a single slot is the same order of magnitude.

Evidently, splitting the spectrum on smaller fragments results in better rate distribution since this is interference limited scenario. Thus, using smaller spectrum fragments and randomization in the femto tier efficiently mitigates the inter/intra tier interference in congested network. Using smaller spectrum fragments also results in more service denied users. However, by rescheduling the dropped users in subsequent time slots, the system is able to guarantee higher data rates per user. Fig. 2c shows the same results for higher number of FBSs, i.e. $N_f = 5000$ which corresponds to lightly loaded network since the number of available PRBs is significantly higher than the overall number of users. As the results suggest, in this case, the corresponding scenario is capacity limited and besides guarantying better rate distribution and higher rates per user, it also shows that in lightly loaded networks it is better to fully reuse all available system bandwidth in the femto tier. These two limiting situations (i.e. congested vs. lightly loaded networks) show that, from network operator perspective, high data rates can be guaranteed with very simple RRM and resource allocation strategies for the femto tier.

The rate distribution when the average number of FBSs is $N_f = 1000$ is shown in Fig. 2b. The results illustrate that there is a trade-off between attaining high data rates per user and the percentage of users that are guaranteed to achieve those rates. If the operator targets high data rate for small percentage of users, then the femto tier should use larger spectrum fragments. However, if the operator wants to guarantee a predefined, lower data rate to higher number of users, the femto tier should use smaller spectrum fragments. Thus, it can be concluded that the femto tier can dynamically adjust the spectrum fragment size, depending on the network state. That leads to optimized spectrum allocation and sharing. To achieve this, the femto tier needs coordination with the macro tier to get information about the current traffic state of the network, which can be easily achieved using the standard network interfaces, such as X2 interface [15].

Fig. 3 shows a comparison of the performance of user association strategies for different association metrics in terms of the rate distribution in congested network. The results are shown for the case when the femto tier uses the smallest spectrum fragment (of only 3 PRBs) since, as shown on Fig. 2a, such spectrum fragmentation provides best performance in congested networks. The two-step, macro-to-femto offloading scheme significantly outperforms all other user-to-BS association strategies around 1Mbps for this scenario. These observations suggest that, in congested networks, when there is a high amount of scheduled traffic in the downlink in a single time slot, the two-step offloading scheme provides better utilization of the available spectrum resources at the femto tier. This reflects in higher spectrum efficiency. However, the improvement of the two-step offloading over the other association strategies vanishes for higher data rates due to the fact that the LTE air interface is channelized system with limited physical resources.

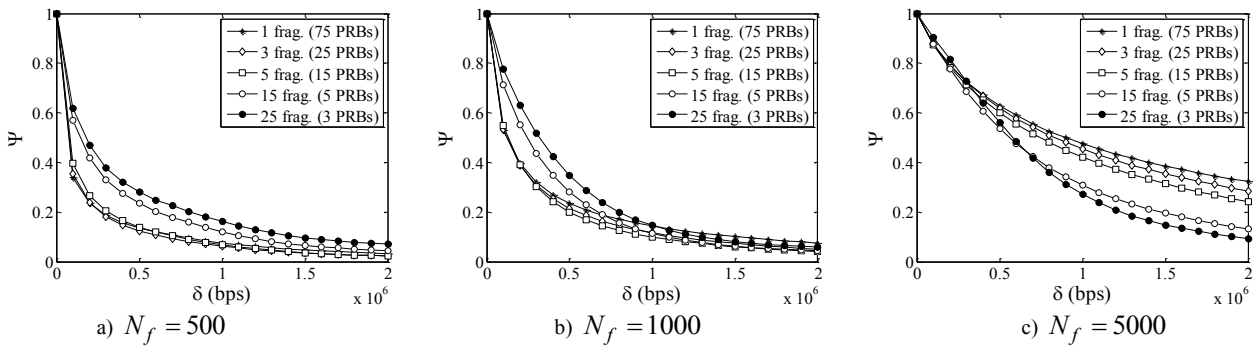

Fig. 2. Rate distribution of the system for different sizes of spectrum fragments used by the femto tier (nearest BS association)

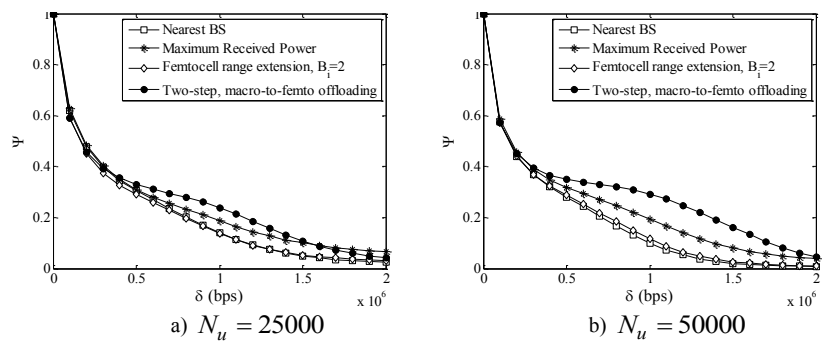

Fig. 3. Rate distribution of the system for different user association strategies

The rate distribution for all users with $\delta = 1 Mbps$ depending on the embedded access control on the femto tier and different percentage of closed/hybrid access FBSs is shown in Fig. 4. Open access allows the femto-tier resources to be utilized by all users maximizing the network's resource utilization. Therefore, we can consider open access as the limit for the network capacity as well as for the rate distribution in this case. Exclusive access to a closed access FBSs for its subscribers and rejection of non-subscribers' requests hugely reduces the utilization of the femto-tier resources and the average rate per user. Retaining the preferential access for the subscribers, hybrid access approaches the capacity of open access scenario keeping the average rate per user high. Hence, hybrid access control emerges as the best solution for a further implementation of access control on cellular FBSs.

## VI. CONCLUSIONS

This paper analyzes the spectrum resource allocation, sharing and utilization efficiency in two-tier, heterogeneous networks with macro and additional, uncoordinated, femto tier. The femto tier uses simple spectrum fragmentation and random fragment allocation to determine the operating resources. The results show that the rate distribution of the system depends on the fragment size for varying network conditions and suggests that the femto tier can dynamically adjust it. The paper also proposes novel, two-step, macro-to-femto offloading scheme for user-to-BS association for efficient utilization of the femto tier resources. The scheme is proven to provide better rate distribution compared to existing user association strategies.

From operators' perspective, the results can be used for intelligent RRM, where the fragment size can be dynamically updated according to the congestion in the network by providing loose coordination with the macro tier.

Introducing access control on the femto tier improves the overall network performances and targeting of specific user groups with different requirements. Operators can guarantee minimal average rates and preferential access for subscribers, offering an opportunity for implementation of newer and richer services. Future work can be done in the practical implementation of the femtocells as well as the business models of their integration into future cellular networks.

ACKNOWLEDGMENT

This work was supported by the Public Diplomacy Division of NATO in the framework of Science for Peace through the SfP-984409 ORCA project.

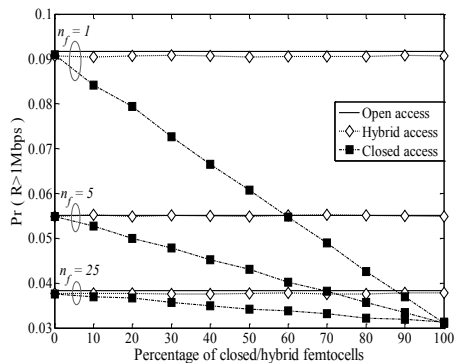

Fig. 4. Rate distribution for different access control mechanisms and varying fragment size ($N_u = 10000, N_f = 600$)